\documentclass{eas}
\usepackage{graphicx}
\begin{document}

\title{Rapidly Star-forming Galaxies At High Redshifts} 
\author{Romeel Dav\'e}\address{University of Arizona, 933 N. Cherry Ave., Tucson, AZ 85721 USA}

 \begin{abstract}
{\it Herschel} has opened new windows into studying the evolution of
rapidly star-forming galaxies out to high redshifts.  Today's massive
starbursts are characterized by star formation rates (SFRs) of $\sim 100+
M_\odot/$yr and display a chaotic morphology and nucleated star formation
indicative of a major merger.  At $z\sim 2$, galaxies of similar mass and
SFR are characterized by ordered rotation and distributed star formation.
The emerging cold accretion paradigm provides an intuitive understanding
for such differences.  In it, halo accretion rates govern the supply of
gas into star-forming regions, modulated by strong outflows.  The high
accretion rates at high-$z$ drive more rapid star formation, while
also making disks thicker and clumpier; the clumps are expected to be
short-lived in the presence of strong galactic outflows as observed.
Hence equivalently rapid star-formers at high redshift are not analogous
to local merger-driven starbursts, but rather to local disks with highly
enhanced accretion rates.
\end{abstract}
\maketitle
\section{Introduction}

{\it Herschel} has revolutionized far-infrared studies of the interstellar
medium (ISM) in rapidly star-forming galaxies at high redshifts.  Recent
observations have
accurately characterizing the peak of their dust-reprocessed emission, and
in some cases even have detected key ISM cooling lines.  This has resulted in
significantly more robust determinations of galaxy star formation rates
and ISM physical conditions, thereby lending crucial insights into the
physics of high-$z$ star formation.

A key result known prior to, but whose robustness was greatly increased
by, {\it Herschel} is that the luminosity density in the infrared evolves
quite strongly with redshift.  For instance, \cite{gru10} determined
that the IR luminosity density evolves as $(1+z)^{3.8}$ up to $z=1$
and continues to increase to earlier epochs, and they attributed this
increase to the prevalance of starburst galaxies at early epochs.
Yet the increase is reflective of an overall evolution in the galaxy
population rather than in a particular class of systems: At any given
stellar mass, the star formation rate of a galaxy at $z=2$ is more
than an order of magnitude higher than one at $z=0$~(\cite{nor10}).
Morphological studies of high-$z$ galaxies generally show a lumpy or
chaotic morphology, which if seen in a galaxy today would surely be
classified as a merger event.  These and related observations are often
taken as strong support of a scenario in which merger-induce starbursts
are more common at higher redshifts, and become largely responsible for
driving galaxy evolution at early epochs.  This scenario meshes well with
the notion of hierarchical structure formation in which small systems
form first and assemble through mergers into larger systems.

In contrast, recent hydrodynamical simulations of cosmological galaxy
formation consistently suggest that mergers are sub-dominant in fueling
galaxy growth (\cite{mur01,ker05,guo08,vdv10}).  Instead, these models
generically predict that gas feeding primarily occurs through cold, smooth
streams (\cite{ker05,dek09,ker09}), and while those streams also carry
in galaxies, major mergers that would induce strong starbursts are rare.
This so-called ``cold accretion" is predicted to be particularly strong
at early times when cosmic star formation peaks, and so is expected to
drive galaxy growth in the epoch now being probed by {\it Herschel} and
other telescopes.  The natural question then arises, are the properties
of high-$z$ galaxies now being detected in accord with expectations from
this scenario?  In particular, how does one explain the rapid increase
with redshift in star formation rates and disturbed morphologies within
the cold accretion paradigm?

\section{Gas Fueling}

Far from galaxies, the motion of baryons is largely governed by the
mass-dominant dark matter.  To a reasonable approximation, the accretion
rate of baryons into halos is then expected to be given by the accretion
rate of dark matter times the cosmic baryon fraction.  Simulations confirm
this, and have shown that it is largely insensitive to feedback processes
(\cite{vdv10}).

The accretion rate onto dark matter halos can be accurately characterized
as a function of mass and redshift assuming a currently-favored
$\Lambda$CDM cosmology.  \cite{nei06} computed halo accretion rates both
from analytic Press-Schecter based arguments as well as from dark matter
(i.e. N-body) simulations, and showed that the baryon halo accretion
rate is well-described by
\begin{equation}\label{eq:acc}
\dot{M}_{\rm in} = 6.6 M_{12}^{1.15} (1+z)^{2.25} f_{0.165}\; M_\odot{\rm yr}^{-1}
\end{equation}
where $M_{12}$ is the halo mass in units of $10^{12} M_\odot$, and 
$f_{0.165}$ is the cosmic baryon fraction in units of the WMAP-concordant
value $f_b=0.165$ (\cite{dek09}).

This equation implies that at $z=0$, a Milky Way-sized galaxy's halo
($10^{12}M_\odot$) will accrete $\sim 7 M_\odot/$yr of gas into its
virial radius.  The star formation rate in the Milky Way is $\sim 1-3
M_\odot$/yr (see \cite{rob10} and references therein), indicating that,
as one might expect, only some of the accretion into the halo's virial
radius actually forms into stars.  We will argue in \S\ref{sec:feedback}
that this is indicative of ejective feedback, but for now let us focus
on the redshift evolution.

A similar-sized halo at $z=2$ will, by equation~\ref{eq:acc}, accrete
at $\approx80 M_\odot$/yr.  Under the simple assumption that a constant
fraction of halo accretion ends up forming into stars, it is immediately
seen that the star formation rate of a given mass galaxy at $z=2$ should
be more than an order of magnitude higher than at $z=0$.  The observed
evolution of the specific star formation rate (sSFR) is indeed consistent
with such an increase out to $z\sim 2$~(\cite{noe07,dad07})

\begin{figure}\label{fig:ssfrevol}
  \includegraphics[width=15cm]{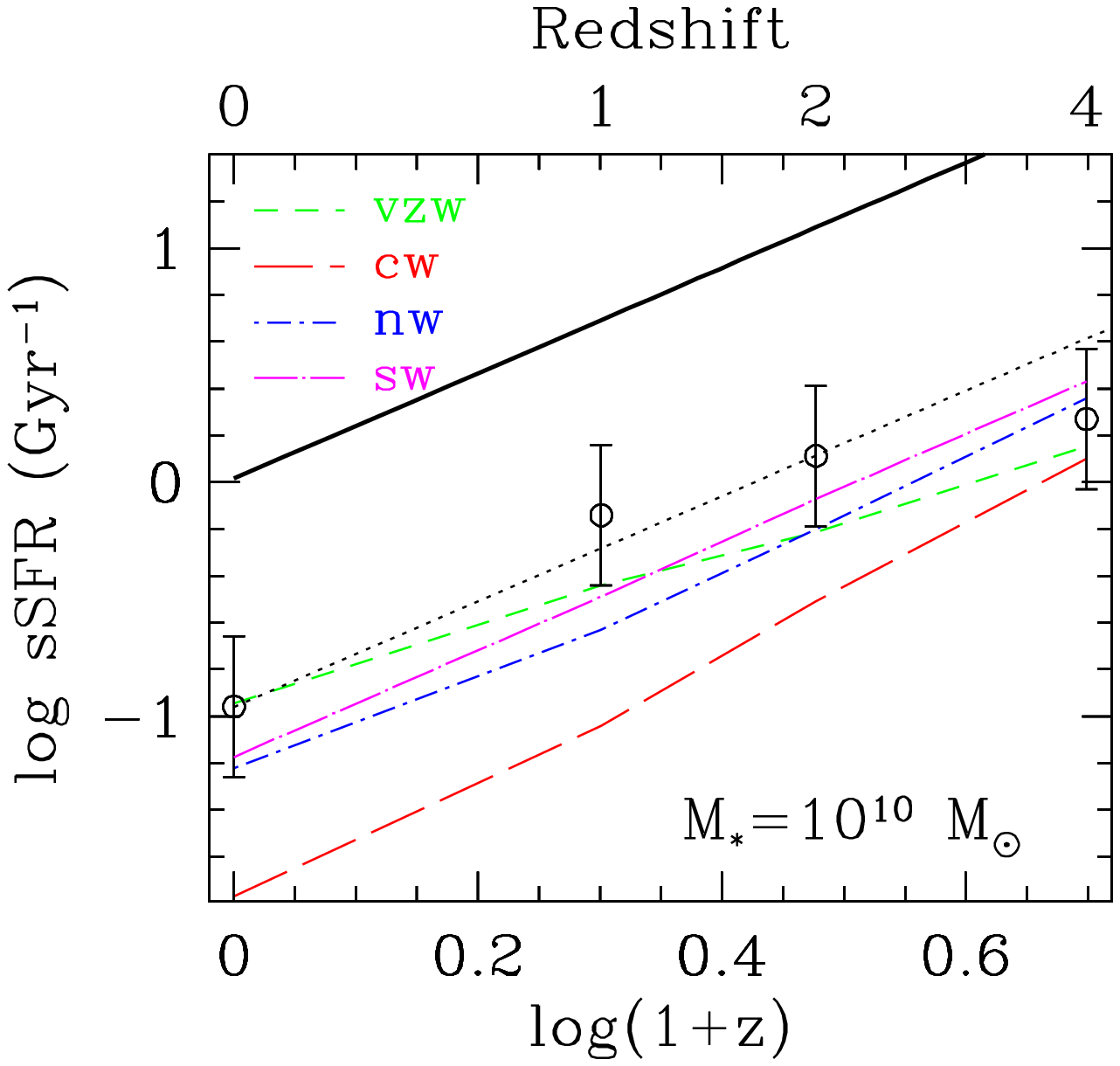}
  \vskip -4.2in

  \caption{Comparison of models to data for the evolution of the
sSFR of a galaxy with $10^{10}M_\odot$ of stars.  Observations, shown
as the circles with an approximate $1\sigma$ spread of $\pm0.3$~dex,
are shown at $z=0,1$ (\cite{elb07}) at $z=2$ (\cite{dad07}) corrected
downwards by a factor of 1.5 as suggested by {\it Herschel} data
(\cite{nor10}), and at $z=4$ from \cite{sta09}.  The model lines of
various colours/line types show results from cosmological hydrodynamic
simulations (\cite{dav11}), with various prescriptions for galactic
outflows, including a model without outflows (nw).  The thick
solid line shows the prediction from equation~\ref{eq:acc} assuming a
halo mass of $2\times 10^{11}M_\odot$, while the dotted line shows a
similar redshift scaling normalized to the observed $z=0$ sSFR.}
\end{figure}

Hydrodynamic simulations that attempt to directly model
the growth of galaxies are also consistent with this scenario.
Figure~\ref{fig:ssfrevol} shows the evolution of the sSFR as a function of
$\log({1+z})$, comparing expectations from the cold accretion paradigm,
predictions from hydrodynamic simulations, and observations.  The thick
solid line shows the power-law relation from equation~\ref{eq:acc}, and
the dotted line shows a similar scaling assuming a constant fraction of
star formation relative to halo accretion, normalized to the observed
$z=0$ data.  Observations are shown as the circles, scaled to a fiducial
stellar mass of $10^{10}M_\odot$ (see \cite{dav11}).  In general, the
observed evolution of the sSFR nicely follows equation~\ref{eq:acc}.
Also shown are results from cosmological hydrodynamic simulations
(\cite{opp10},\cite{dav11}) with various prescriptions for galactic
outflows.  In each case, the evolutionary trend broadly follows that
expected for cold accretion.  The trend itself is relatively insensitive
to feedback prescription (\cite{dav08,dut10}), but the overall amplitude
can be affected in extreme feedback models.

Hence the cold accretion paradigm naturally explains the overall reduction
in cosmic star formation rates at a given mass as a direct consequence
of dropping hierarchical mass accretion rates.  The key implication
for understanding galaxy evolution is that galaxies with an order of
magnitude higher SFRs at high-$z$ should not be seen as analogs of
local starbursts, but rather analogs of local disks being smoothly fed
at much higher accretion rates.  Just because $z\sim 2$ galaxies have
high star formation rates does not make them starbursts; they are simply
processing gas as supplied, and hence are ``supply-limited" just as local
disks today.  The gas supply can be strongly tempered by feedback as we
discuss next, but the simple ansatz of a roughly constant fraction of the
virial radius accretion forming into stars broadly matches observations.

\section{The Role of Feedback}\label{sec:feedback}

In simulations without galactic outflows, the accretion rate onto
a galaxy is only slightly smaller than that at its virial radius
(\cite{ker09,vdv10,cev10}) in cold accretion-dominated halos (i.e. $M_{\rm
halo}<\sim 10^{12}M_\odot$).  This is because cold streams operate
to efficiently channel gas down filaments connecting to large scale
structure, and the enhanced density within the filaments prevents the
formation of an accretion shock along its path.  It is therefore difficult
to stop rapid accretion onto a galaxy unless the halo is sufficiently
large so as to disrupt the cold streams, and perhaps not even then
(\cite{ker09b}).

If most of the gas accreting into the virial radius reaches near the disk,
one would expect in a steady-state situation that the star formation
rate would be comparable to the accretion rate.  This contradicts
observations of, e.g., the Milky Way, where the SFR is only a fraction of
the accretion rate.  Furthermore, the baryon fraction in stars relative to
the overall baryon fraction expected in halos is small, in fact peaking
around Milky Way-sized objects at $\sim 25\%$ and dropping off to either
higher or lower masses (\cite{mcg10,dai10}).  Hence something must be
preventing that accreted gas from forming into stars.  Given that galaxy
gas fractions are seen to be dropping with time (\cite{tac10}), the gas
cannot be piling up in the disk unused (barring it being in some hidden
form, which appears unlikely).  Hence an inevitable conclusion of the
cold accretion paradigm is that galactic outflows, i.e. kinetic removal
of gas from the star-forming regions, must be prevalent.

Observations are amassing that outflows are ubiquitous in high-redshift
galaxies (e.g. \cite{ste04,wei09,ste10}).  Such outflows have also been
invoked in models to enrich the intergalactic medium (e.g. \cite{opp06})
and modulate the growth of stellar mass (e.g. \cite{opp10,dav11}) and
metallicity (e.g. \cite{fin08}).  Perhaps the most surprising aspect of
these models is that the amount of material ejected must significantly
exceed the amount forming into stars (\cite{opp08}).  In the cold
accretion paradigm, this makes sense because the star formation rates
are a small fraction of the accretion rates, and hence the rest must
be ejected.

Overall, cold accretion must work together with outflows if this paradigm
is to yield galaxies as observed.  To zeroeth order, there is a simple
balance between gas accretion at the virial radius supplying gas, and
star formation and outflows removing gas from the ISM.  The evolution
of this equilibrium then governs the observable properties of galaxies
across cosmic time (\cite{dav11}).  In detail, the amount of accretion
onto galaxies does depend on feedback processes, even in the low-mass
regime dominated by cold accretion (\cite{vdv10}).  Furthermore,
winds often recycle back into galaxies, providing another accretion
path that may be dominant at late epochs (\cite{opp10}).  Nevertheless,
this simple scenario is at least broadly consistent with available data,
and can help to place the observed evolution of galaxies out to high
redshifts within a hierarchical context.

\section{Morphology and Kinematics}

The morphology and kinematics of high-$z$ galaxies have drawn considerable
attention, particularly with the emergence of near-infrared integral field
units on large telescopes such as SINFONI and OSIRIS.  The SINS survey
(\cite{for09}) has revealed the particularly interesting trend that
higher-mass galaxies at $z\sim 2$ tend to show more ordered rotation,
despite having clumpy and irregular H$\alpha$ maps that suggest
ongoing interactions (\cite{gen06}).  It is possible that the apparent
rotation could actually be two merging similar-mass galaxies in orbit
(\cite{rob08}), but the prevalance of rotation signatures together with
the expected rarity of major mergers caught exactly at such a stage
means that such an explanation is unlikely for most of these objects.

Reconciling the clumpy morphology and ordered kinematics of $z\sim 2$
galaxies has provoked a number of simulators to re-examine disk galaxy
formation tailored to high redshifts.  \cite{bou08} showed that a disk
evolving in isolation under conditions as observed at high-$z$ would
fragment quickly, producing a clumpy morphology with ordered kinematics
as observed.  This requires that the initial bulge component be small,
which \cite{bou09} argued cannot be the case if mergers (which are
believed to grow bulges) dominate early galaxy growth.  The clumps
form from self-gravitational instabilities within the disk, without
associated dark matter halos, and can migrate inwards to grow a bulge
secularly (\cite{cev10}).  \cite{bou10} showed using AMR hydrodynamic
simulations gravity-driven turbulence feedback can produce thick disks
with substantial velocity dispersion, in agreement with observations.
Hence overall, the secular formation of clumpy disks seems to be at least
plausible given the conditions (particularly the high gas fractions)
of high-$z$ galaxies.

However, the issue of clump growth and migration has elicited some
concerns.  If the clumps are long-lived and sink to the middle, and are
a ubiquitous feature of high-$z$ galaxies, then it becomes difficult
to explain the formation of late-type galaxies today; this is the
longstanding angular momentum problem (e.g. \cite{ste99}).  Furthermore,
we argued earlier that the SFR cannot be comparable to the accretion rate
over long timescales, as happens in these models (e.g. \cite{cev10}),
since this would significantly overproduce the fraction of baryons
in stars (i.e. the longstanding overcooling problem).  Indeed, the
fragmentation of disks in models without significant ISM feedback is
a well-known result that has generally been regarded as a catastrophic
failure by those aiming to simulate present-day disks (e.g. \cite{rob04}).
Hence either the clumpy disk interpretation of the observations is flawed,
or else some physical ingredient is missing in these models.

The missing ingredient is likely to be galactic outflows.  \cite{gen11}
studied the formation of a high-$z$ disk galaxy including outflows.
The model for outflows was heuristic and parameterized, but it followed
an outflow model demonstrated to broadly match a range of galaxy and
intergalactic medium properties.  The key new finding is that while
marginally-stable clumps form as found by others, the mass loss owing to
strong outflows enables clump disruption on short timescales, well before
they are able to sink to the middle and grow a bulge.  Their resulting
galaxy showed a significantly smaller fraction of baryons in stars,
in better accord with data.  The properties of these simulated clumps
are in good agreement with SINFONI observations of individual clumps
by \cite{genz11}, which also show strong outflows with mass loss rates
significantly exceeding their SFR, and modest turbulence likely driven
mostly by gravity.

The implications for disrupting clumps are numerous and, in general,
favorable towards forming disk galaxies as observed.  In particular,
it has long been suggested that the angular momentum problem may be
solved if outflows preferentially remove low angular momentum gas.
The specific angular momentum distributions within dwarf galaxies today
show a marked deficit of low angular momentum gas relative to expectations
from a simple disk collapse model (\cite{vdb01}).  Indeed, \cite{gov09}
was able to form a bulgeless dwarf galaxy by including strong feedback
which they argued removed the low angular momentum gas, although it
remains to be seen if the same mechanism can operate in a larger galaxy.
The disruption of clumps by outflows suggests a slight modification to
the canonical interpretation: It is not that low angular momentum gas is
preferentially removed, but rather that gas is expelled from clumps before
it would have otherwise lost its angular momentum and grown the bulge.
This alleviates the need for outflows to originate near the centers of
growing galaxies, and can instead be driven from where star formation
is occuring.

\section{Summary}

The evolution of star-forming galaxies from $z\sim 2\rightarrow 0$
has yielded several paradoxical results in terms of the importance
of mergers in driving their evolution.  The disturbed morphologies
and high star formation rates at $z\sim 2$ are often analogised with
mergers seen today.  Conversely, the kinematics show ordered rotation,
and detailed analyses of lensed galaxies' infrared SED suggest that they
are more similar to local quiescent disks than starbursts (\cite{rex10}),
disfavoring the merger hypothesis.

The seemingly conflicting results can be reconciled within the cold
accretion paradigm for galaxy growth.  The high star formation rates
owe to increased accretion rates at high-$z$, which result in secular
morphological features such as clumps that are not seen in local disks.
Most of these systems are not therefore undergoing major mergers; at
$z\sim 2$, such mergers probably give rise to the brightest sub-millimeter
galaxies, i.e.  hyper-LIRGs rather than ULIRGs.

Because it is difficult to prevent dense filamentary cold streams from
reaching the disk, the cold accretion scenario must necessarily be
accompanied by a mechanism to remove a substantial amount of gas from
the star-forming regions, such as galactic outflows.  This is motivated
by observations that star formation rates are substantially smaller than
halo accretion rates in a $\Lambda$CDM universe, and that the fraction
of baryons forming into stars within halos is small.  Outflows can also
have significant impact on the internal structure of high-$z$ galaxies,
including causing more rapid disruption of clumps and preventing the
ubiquitous formation of large bulges.  Interestingly, both simulations
and observations seem to suggest that the turbulence of high-$z$ disks
is {\it not} a result of feedback but rather is powered by gravitational
potential energy, suggesting that outflows escape without depositing
much of their energy into their immediate surroundings.

The interplay between accretion, star formation, and outflows remains
poorly understood.  Simulations are just beginning to explore the
connections that span a range of scales from parsecs to megaparsecs.
Observations of high-$z$ galaxies are critical for pinning down the
key physical processes driving galaxy evolution, and {\it Herschel}
has played, and will continue to play, a central role in that effort.

 \acknowledgements
The author wishes to thank Ben Oppenheimer and Kristian Finlator
for extensive discussions and assistance, and Avishai Dekel, Natasha
F\"orster Schreiber, Reinhard Genzel, Neal Katz, Du\v{s}an Kere\v{s},
Linda Tacconi, and David Weinberg for helpful conversations, along with
the organizers for an excellent conference in a great location.


\end{document}